\documentclass[twocolumn,showkeys,preprintnumbers,amsmath,amssymb]{revtex4}
\usepackage{graphicx}
\usepackage{epstopdf}
\usepackage{bm}
\usepackage{graphicx}

\begin{document}

\preprint{APS/v0.0 23/10/2008}

\title{Measuring the Hierarchy of Feedforward Networks}
\author{Bernat Corominas-Murtra$^{1,2}$, Carlos Rodr\'iguez-Caso$^{1,2}$ , Joaquin Go\~ni$^3$ and Ricard V. Sol\'e$^{1,2,4}$}

\affiliation{$^1$ ICREA-Complex Systems Lab, Universitat Pompeu Fabra (GRIB-PRBB). Dr Aiguader 88, 08003 Barcelona, Spain\\
$^2$ Institut de Biologia Evolutiva. CSIC-UPF. Passeig Maritim de la Barceloneta, 37-49, 08003 Barcelona, Spain\\
$^3$ Neurosciences Department, Center for Applied Medical Research. 
University of Navarra. Pamplona, Spain
$^4$Santa Fe Institute, 1399 Hyde Park Road, New Mexico 87501, USA}
\thanks{Author for Correspondence:\\ bernat.corominas@upf.edu\\
carlos.rodriguez@upf.edu}



\begin{abstract}
In  this paper we explore the concept of hierarchy  as a quantifiable descriptor of ordered structures,
departing from the definition of three conditions to be satisfied for a hierarchical structure:
 {\em  order},  {\em  predictability}  and  {\em  pyramidal  structure}. According to these principles 
 we define a hierarchical index taking concepts from graph and information theory. This estimator allows to 
 quantify the hierarchical  character of any system susceptible to be abstracted in a feedforward causal graph, i.e.,
 a directed acyclic graph defined in a single connected structure. Our hierarchical  index is a balance between
 this predictability and pyramidal condition by the definition of two entropies: one attending the onward flow
 and other for the backward reversion.
 We show how this index allows to identify hierarchical,  anti-hierarchical and non hierarchical structures.
 Our formalism  reveals that departing from the defined conditions for a hierarchical structure, feedforward trees
 and the inverted tree graphs emerge as the only causal structures of maximal hierarchical and anti-hierarchical 
systems, respectively. Conversely, null  values of the hierarchical index are attributed to  a number of different
 configuration networks; from linear chains, due to their lack of pyramid structure, to full-connected feedforward 
graphs where the diversity of onward pathways is canceled by  the uncertainty (lack  of predictability) when going backwards. 
 Some illustrative examples are provided for the distinction among these three types of hierarchical causal graphs.

\end{abstract}

\keywords{Directed Acyclic Graphs, Information Theory, Order Theory, Hierarchy}
\maketitle


{\bf The idea of hierarchy  has been largely attributed to a disparate number of systems and, 
although easily perceived, its quantification is not a trivial issue. In this work we quantify
 the hierarchy of a given causal structure with a feedforward structure.  Starting with the 
representation of a system of causal relations as a graph, we define a non heuristic measure 
of hierarchy having strong grounds on the principles of information theory. We depart from the definition
of the conditions for a system to be considered perfectly hierarchical: a pyramidal
 structure with a completely predictable reversion of the causal flow. In this context, a hierarchy 
index is defined by weighting how far is a given feedforward structure from these conditions. As we shall 
see, structures that fully satisfy this property belong to a special class of trees. Our 
estimator allows to establish a quantitative criterion for the definition of hierarchic,  
non-hierarchic and anti-hierarchic networks.}

\section{Introduction}
\label{H}
 
The existence of some sort of hierarchical order is an  apparently  widespread feature  of  many  complex  systems, 
including gene \cite{Yu2006} and human brain \cite{Bassett2008,Meunier2009} networks, ecosystems \cite{Hirata1985,Miller2008}, 
social and urban structures \cite{Kolars1974}, the Internet \cite{Vazquez2002} or open-source communities \cite{Valverde2007}. 
The presence of such underlying order in the multiscale 
organization  of complex systems  is a long standing hypothesis \cite{Whyte1969} giving rise to the idea of
hierarchy as a central concept -see also \cite{Huberman1986}. Although usually 
treated only in qualitative terms,  some formal approaches to the problem have been proposed. 
The efforts towards a well-defined quantification of hierarchical order have been 
improving by means of complex networks theory. As a key part of their organization, 
dedicated efforts have been made towards a proper identification of 
hierarchical trends. One outcome of these efforts has been a number
of powerful, heuristic measures \cite{Clauset2008a,Ma2004a,Ravasz2002,
Sales-Pardo2007,Trusina2004,Vazquez2002}.

\begin{figure*}
\begin{center}
\includegraphics[width=14cm]{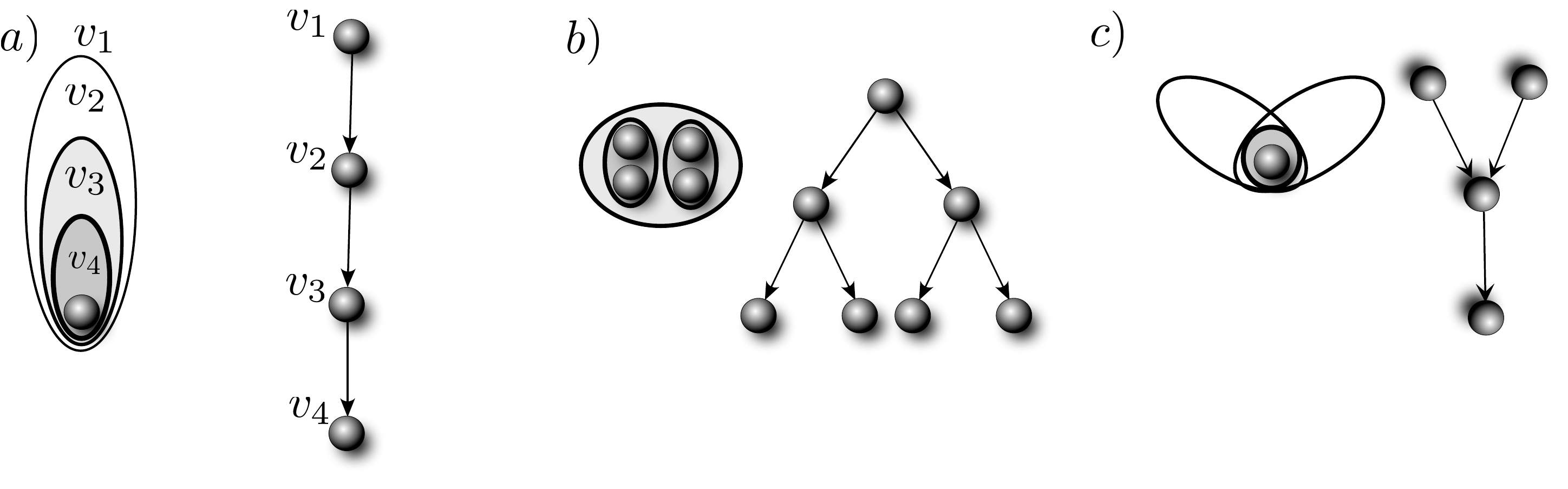}
\caption{Order and hierarchy. a) Total or
  complete order either defined by inclusion displaying \textit{nested
ensembles}, $v_1\subset\ v_2 \subset v_3\subset v_4$ or in terms of
  the order relation $v_1> v_2 > v_3 > v_4$  (left). The direct or immediate relation depicted by
  its \textit{causal  graph} (right).  b) The ideal hierarchy structure assumed
in  this   work represented by its nested organization (left) 
and its causal graph  (right). c) An example of a 
partial ordered defined by inclusion (left) and its respective causal graph (right).}
\label{fig1}
\end{center}
\end{figure*}

Often, a nested organization -formally identical to order, in set-theoretical terms \cite{Kuratowski:1921}- can be identified as the basis of hierarchical order. If we
think in hierarchy  in these terms, we might  agree with Herbert Simon
that  "it is  a  commonplace that  nature  loves hierarchies" -cited in 
\cite{Pattee1973}. Many examples belong to this picture. Within the context of matter organization, 
molecules  are made of atoms, which result from the combination of 
elementary particles, some of which having also internal subunits (as quarks). 
Similarly, the  relation of  characteristic  scales  of organization by  inclusion of  one in another, 
like Chinese  boxes or Matryoshka dolls, has been seen as a sort of hierarchical organization. 
Another relevant example is the case of fractal structures which naturally define 
a hierarchy of self-similar objects. Finally, within the context of taxonomy, spin glasses or optimization theory, 
the use of {\em ultrametricity} has also allowed to define hierarchical order \cite{Rammal:1986}.

Biological hierarchies have also evolved through time as part of a process 
that generates high-level entities out of nesting lower-level ones \cite{Eldredge1985,McShea2001}.  
However, taxonomic classification trees are probably the most obvious 
representation of a hierarchy  based on inclusion. In biology, 
living  beings are  individuals  grouped  in \textit{taxa}
according  their characteristics. Starting from species, a nested 
hierarchy is defined where species belong to genera, which are included within families 
forming orders and so forth. Here, every organism is, in principle, unambiguously classified
and  therefore no  uncertainty  can  be associated  to  the process  of
classification.

Alternatively to  the view of nestedness,  a hierarchical organization
can  be defined  in  a structure  of  causal relations.   Paradigmatic
examples are the flowchart of a company or the chain of command in the
army, where the authority concept  defines a particular case of causal
relation.   Causality   induces  an  asymmetrical   link  between  two
elements, and this asymmetry, either  by inclusion or through any kind
of causal relation, defines an order.  We observe that any circular or
symmetrical  relation between  two  elements violates  the concept  of
order, thereby intuitively loosing its hierarchical nature.  From this
perspective,     any    feedforward     relation     is    potentially
hierarchical. Such  feedforward structures pervade a  diverse range of
phenomena and structures, from phylogenetic trees to river basins.

A common feature of most of the approaches mentioned above is that the
notion of {\em hierarchy} is  basically identified with the concept of
{\em  order}.   Order  is  a  well  defined   concept  in  mathematics
\cite{Kuratowski:1921, Suppes:1960} but, is  order enough to grasp the
intuitive  idea of  hierarchy? Can  we actually  define what  is {\em
  hierarchy}? Quoting Herbert Simon, \cite{Simon:1962}.
\begin{quote}
{\em  (...) a  hierarchical system  -or hierarchy-  can be  defined in
  principle as  a system that is composed  of interrelated subsystems,
  each  of the  latter being  also hierarchic  in structure  until the
  lowest  scale is reached\footnote{We  observe that  this definition,
    although circular,  captures the  idea that a  hierarchical system
    displays a kind of regularity that is repeated along the different
    scales of the system.}.}
\end{quote}
This definition does not provide a clear formalization of hierarchy as
a measurable feature, although it certainly grasps the intuitive idea of hierarchy.
How can such a general measure be defined?  It is reasonable to assume
that we have a hierarchy if  there is no ambiguity (or uncertainty) in
the chain of  command followed for any individual to  the chief in the
flowchart.  We   shall  call   this  feature  the   {\em  definiteness
  condition}.   This is also  valid for  nested structures.  One might
think that this condition is justified by a single chain of command or
in the case of matryoshka  dolls. However, such structures are already
defined   within  order  theory   as  \textit{totally   ordered}  (see
fig. \ref{fig1}a).  Order and hierarchy  are closely related  but they
are not essentially the same.  In  this paper we reserve the word {\em
  hierarchy} to designate a concept that goes beyond the definition of
order.   We argue  that  the difference  stems  from the  fact that  a
hierarchical   structure   must   also   satisfy  a   {\em   pyramidal
  organization} constraint -see  fig.  (\ref{fig1}b).  In other words,
the lower the layer of organization, the larger the number of entities
it contains. But,  what happens when  this pyramid structure is
inverted?   Intuitively,   they   would   not  be   hierarchical   but
anti-hierarchical\footnote{Anti-hierarchy is not  a new concept in the
  field of  complex networks \cite{Trusina2004}. As we  shall see, the
  hierarchical  index is  calculated by  a difference  of  two entropy
  measures and  thus, hierarchical index allows  negative and positive
  values. In this context, terms such as hierarchy, anti-hierarchy and
  non-hierarchy   (or  alternatively   positive,   negative  or   null
  hierarchy) could be  interpreted with the same meaning  as it occurs
  in the  Pearson coefficient where positive  correlation and negative
  correlation   (anti-correlation)    are   opposite   tendencies   in
  contraposition  with the  null tendency  of a  non-correlation.}. In
this  work  we show  how  information  theory  naturally provides  the
suitable  framework to  characterize hierarchy  in  causal structures.
Within this  theoretical apparatus we provide  the rigorous definition
of the hierarchy index for causal  structures and how it is applied in
some   illustrative   examples   establishing  a   distinction   among
hierarchical, non-hierarchical and anti-hierarchical.

\section{Directed graphs, Ordered Graphs and Causal Graphs}
\label{BasicGDef}

In this section we will present the basic theoretical background 
used in this paper, grounded both in order and graph theories. At the end of this section we will formally 
define the \textit{causal graph}, the key concept of this theoretical framework 
where the proposed hierarchy measures are  applied. 

\subsection{Basic concepts of order}

Hierarchy is undoubtedly tied to order. This is why we make
a brief review of order  theory highlighting some features that have
been commonly attributed to hierarchy.  The first task is to define an
ordered pair between two elements $a_k,a_j$  of a given set $A$, to be
written  as  $a_k>a_j$, $\langle  a_k,a_j\rangle$  or,  in a  formally
equivalent way:
\[
\langle a_k,a_j\rangle=\{\{a_k\},\{a_k,a_j\}\}.
\]
This latter  formalization explicitly defines order  from an inclusion
relation \cite{Kuratowski:1921}. This immediately connects  to standard views of hierarchical
systems,  as  we  already mentioned,  in  which  inclusion
relations are considered essential. Having defined an ordered pair, we
define an order relation.  Let $A=\{a_1,...,a_n,...\}$ be a countable,
finite set and ${\cal R}\in A\times  A$ a relation. Such a relation is
an {\em  order relation} -rigorously  speaking, a {\em  strict partial
  order}- if the following condition holds:
\begin{eqnarray}
&i)&\langle a_k, a_k \rangle \notin {\cal R},\nonumber\\ &ii)&(\langle
  a_i,  a_k\rangle \in {\cal  R})\Rightarrow (\langle  a_k, a_i\rangle
  \notin  {\cal  R}),\nonumber\\  &iii)&(\langle a_i,  a_k\rangle  \in
         {\cal  R}   \wedge  \langle   a_k,  a_j  \rangle   \in  {\cal
           R})\Rightarrow   (\langle  a_i,   a_j  \rangle   \in  {\cal
           R}).\nonumber
\end{eqnarray}

We finally define two  subsets of
$A$  from the definition  of order  relation which  will be  useful to
characterize the kind of structures  studied in this paper. The set of
maximal elements of $A$, to be written as $M\subset A$, is defined as:
  \[
 M=\{a_k\in  A:  \nexists a_j\in  A:  \langle a_j,a_k\rangle\in  {\cal
   R}\}.
  \]
Similarly,  the  set  of  minimal  elements, to  be  written  as
$\mu\subset A$ is defined as:
\[
\mu=\{a_k\in  A: \nexists  a_j\in A:  \langle  a_k,a_j\rangle\in {\cal
  R}\}.
\]

\subsection{Basic concepts of Directed Acyclic Graphs}

Let    ${\cal    G}(V,    E)$    be   a    directed    graph,    being
$V=\{v_1,...,v_n\},\;|V|=n$, the  set of nodes,  and $E=\{\langle v_k,
v_i\rangle, ..., \langle v_j, v_l\rangle\}$ the set of arcs -where the
order, $\langle v_k,  v_i\rangle$ implies that there is  an arc in the
following direction:  $v_k\rightarrow v_i$.  Given a  node $v_i\in V$,
the  number of  outgoing links,  to be  written as  $k_{out}(v_i)$, is
called the {\em  out-degree} of $v_i$ and the  number of ingoing links
of  $v_i$  is  called  the   {\em  in-degree}  of  $v_i$,  written  as
$k_{in}(v_i)$. The {\em adjacency matrix} of a given graph ${\cal G}$,
$\mathbf{A}({\cal   G})$   is   defined   as  $A_{ij}({\cal   G})=   1
\leftrightarrow \langle v_i,  v_j\rangle\in E$; and $A_{ij}({\cal G})=
0$ otherwise.   Through the  adjacency matrix, $k_{in}$  and $k_{out}$
are computed as
\begin{equation}
k_{in}(v_i)=\sum_{j\leq                                  n}A_{ji}({\cal
  G});\;\;\;\;k_{out}(v_i)=\sum_{j\leq n}A_{ij}({\cal G}).
\label{kin}
\end{equation}
Furthermore, we will  use the known relation between  the $k$-th power
of the  adjacency matrix and the  number of paths of  length $k$ going
from a given node $v_i$ to a given node $v_j$ Specifically,
\begin{equation}
(\mathbf{A}^k({\cal              G}))_{ij}=(\overbrace{\mathbf{A}({\cal
      G})\times...       \times     \mathbf{A}({\cal     G})}^{k\;{\rm
      times}})_{ij}
      \label{Ak}
\end{equation}
is the  number of paths  of length $k$  going from node $v_i$  to node
$v_j$ \cite{Gross:1998}.

It is said that $v_i$  {\em dominates} $v_k$ if $\langle v_i,v_k\rangle\in
E$.   A  {\em feed-forward}  or  {\em  directed  acyclic graph} (DAG) is  a
directed graph characterized  by the absence of cycles:  If there is a
{\em  directed path}  from $v_i$  to $v_k$  (i.e., there  is  a finite
sequence  $\langle v_i,  v_j\rangle,  \langle v_j,  v_l\rangle,\langle
v_l, v_s\rangle, ...,  \langle v_m, v_k\rangle \in E$)  then, there is
no  directed  path  from   $v_k$  to  $v_i$.  Conversely,  the  matrix
$\mathbf{A}^T({\cal  G})$  depicts  a  DAG with  the  same  underlying
structure  but having  all  the  arrows (and  thus,  the causal  flow)
inverted.  The {\em underlying graph} of a given DAG ${\cal G}$, to be
written as  ${\cal G}_u$, is the undirected  graph ${\cal G}^u(V,E^u)$
obtained by  substituting all arcs  of $E$, $\langle  v_i, v_k\rangle,
\langle  v_j, v_s\rangle,....$ by  edges giving  the set  $E^u=\{ v_i,
v_k\},  \{ v_j,  v_s\},....$. A  DAG  ${\cal G}$  is said  to be  {\em
  connected}  if for  any pair  of nodes  combination $v_i,  v_l\in V$
there is a finite sequence of pairs having the following structure
\[  
  \{v_i, v_k\}, \{v_k, v_j\},...,\{v_m,v_s\},\{v_s,v_l\},
\]
being  $  \{v_i,  v_k\},  \{v_k,  v_j\},...,\{v_m,v_s\},\{v_s,v_l\}\in
E_u$.

Given the acyclic  nature of a DAG,  one can find a  finite value $L({\cal
  G})$ as follows:
\begin{equation}
L({\cal    G})=\max\{k:(\exists   v_i,    v_j\in   V:(\mathbf{A}^k({\cal
  G}))_{ij}\neq 0)\}.
\label{K}
\end{equation}
It is easy to see that $L({\cal G})$ is the length of the longest path
of the  graph. 

Borrowing concepts from order theory \cite{Suppes:1960}, we define the
following set:
\begin{equation}
M=\{v_i\in V:k_{in}(v_i)=0\},
\label{MaximalSet}
\end{equation}
to be  named the set  of {\em maximal  nodes} of ${\cal G}$,  by which
$|M|=m$.  Additionally, one can define the set of nodes $\mu$ as
\begin{equation}
\mu=\{v_i\in V:k_{out}(v_i)=0\}
\label{minimalSet}
\end{equation}
to be referred as the set of {\em minimal nodes} of ${\cal G}$.

The set  of all  paths $\pi_1,...,\pi_s$, $s\geq  |E|$, from $M$  to a
given node $v_i\in \mu$ is indicated as $\Pi_{M\mu}({\cal G})$.  Given
a node $v_i\in \mu$, the set of all paths from $M$ to $v_i$ is written
as  
\[
\Pi_{M\mu}(v_i)\subseteq \Pi_{M\mu}({\cal G}).
\]
   Furthermore, we
will define the  set $v(\pi_k)$ as the set  of all nodes participating
in  this   path,  except  the   maximal  one.   Conversely,   the  set
$\tilde{v}(\pi_k)$ is the set of all nodes participating on this path,
except the minimal  one.  Attending to the node  relations depicted by
the  arrows,  and due  to  the acyclic  property,  at  least one  node
ordering  can be defined,  establishing a  natural link  between order
theory and  DAGs.  This  order is achieved  by labeling all  the nodes
with  sequential natural  numbers and  obtaining a  configuration such
that:
\begin{equation}
(\forall \langle v_i, v_j \rangle \in E)(i<j).
\label{i>j}
\end{equation}
The existence of this  labeling connects order relations with directed
acyclic graphs.

Finally, throughout this paper we reserve the word {\em tree} to  refer to  those graphs
 where  all nodes excluding the maximal  one have $k_{in}=1$ and all nodes except the minimal 
 ones display $k_{out}>1$. Therefore, we distinguish between chains (all nodes 
with $k_{out}=1$ excluding the minimal one) and trees.

\subsection{Causal Graphs}

In  a  causal graph  we  only  consider  immediate relations between elements i.e.  two
elements are  causally related if there exists  just one cause-effect
event  relating them.  We explicitly  neglect those  relations between
nodes which can only be derived by transitivity. A causal relation can
be    illustrated by genetic inheritance in a genealogy. Offspring's characters come from its parents and indirectly
from  its grandparents. Therefore,  no direct  causal relation  can be
defined between  grandparents and grandsons. However, it  is true that
grandparents indirectly determine the  characters of grandsons, due to
the transitive nature  of the genetic relations. 

In  this work we will
restrict the use of the term  {\em causal relation} to refer to direct
relations such  as direct parent-sons relations, as  described in the
above example. A causal graph ${\cal  G}(V,E)$ is a directed graph where $V$
are  the elements of  a set  (the members  of a  family, in  the above
described example) and $E$ are  the {\em causal relations} that can be
defined between the members of $V$. In this work, we restrict the term
{\em causal  graphs} to graphs  being 
{\em acyclic}  (i.e., DAGs) and  
 {\em  connected}. 
 The  former property  avoids
conflicts in the  definition of the causal flow.  The latter property assumes that two non-connected causal structures
have no  relation among them, and therefore must be considered as two
independent systems. Hereafter, we will refer to the set of paths $\Pi_{M\mu}({\cal G})$ as the set of {\em causal paths}.

\section{The conceptual background of hierarchy}
\label{section:Conceptual}

In  this section we  propose the  basis for  a rigorous  evaluation of
hierarchy. We  begin by defining the  features of what  we consider as
the  {\em  perfect  hierarchical  structure}. As will be shown below, our
proposed  definition of  hierarchy matches  with an  ordered structure
with special  features, thereby making an  explicit difference between
order and hierarchy. Therefore, we reserve the term hierarchy to refer
to a special class of order. 
Within  the  framework  of  graph  theory,  the required  conditions
naturally match those displayed by a tree-like feedforward graph.  Then, as  we shall
see, the estimator we  propose identifies the feed-forward tree topology
as perfectly hierarchical. The main  point  of  the section  is  devoted to  the
definition  of a  quantitative  estimator of  hierarchy  based on  two
entropy measures  that captures the  intuitive ideas described  in the
introductory  section:  the  definiteness and  pyramidal  organization
condition.  We  stress that the forthcoming formalism  applies only to
the class of {\em causal graphs}.
 
\subsection{The starting point: Defining the perfect Hierarchy}

We are going to refer to a system as {\em perfectly  hierarchical} if it satisfies the following conditions.
Let  us consider  a system  depicted by  a {\em  causal  graph} ${\cal
  G}(V,E)$.  We say  that  this  graph ${\cal  G}$  will be  perfectly
hierarchical if the following two conditions hold:
\begin{enumerate}
\item
{\em  Definiteness condition.-} For  every element  $v_k\in V\setminus M$
there is only one element $v_i\in V$, $v_i\neq v_k$ such that
$\langle v_i,v_k\rangle\in E$. A straightforward consequence of this condition is that
\[
m=1.
\]
\item
{\em     Pyramidal    condition.-}     There     is    a     partition
$W=\{\omega_1,...,\omega_m\}$ of the set $V$, i.e.,
\[
V=\bigcup_{W}\omega      _i;\;\;\;\;\;\forall     \omega_i,\omega_k\in
W,\omega_i\bigcap\omega_k= \varnothing
\] 
by which:
\[
(\forall     \langle     v_i,     v_{\ell}\rangle\in    E)     (v_i\in
\omega_j)\Rightarrow (k_{out}(v_i)>1)\wedge(v_{\ell}\in \omega_{j+1}).
\]
A direct consequence of the above is that
\[
|\omega_1|< |\omega_2|<...< |\omega_{|W|}|,
\]
which reflects the pyramidal structure of the graph.
\end{enumerate}
A measure of hierarchy must  properly weight the deviations of the
studied  graph from  the above  requirements.  One  could  add another
condition, by imposing that every  node in a given layer dominates the
same  number of  nodes contained  in the next downstream layer.  In formal
terms, this implies that, in  addition to the above conditions, we can
add a third one, namely:

3. {\em Symmetry condition}. It is established by means of
\[
(\forall    v_i,    v_{\ell}\in   \omega_j)(k_{out}(v_i)=k)\Rightarrow
(k_{out}(v_{\ell})=k).
\]
Which actually corresponds to a so-called complete $k$-ary tree (Gross and Yellen, 1999). 
Therefore,  in those cases where symmetry  is considered  as an inherent feature of
ideal hierarchy, deviations  from symmetrical configurations must also
be taken into account in our quantitative approximation.

\begin{figure}
\begin{center}
\includegraphics[width=7cm]{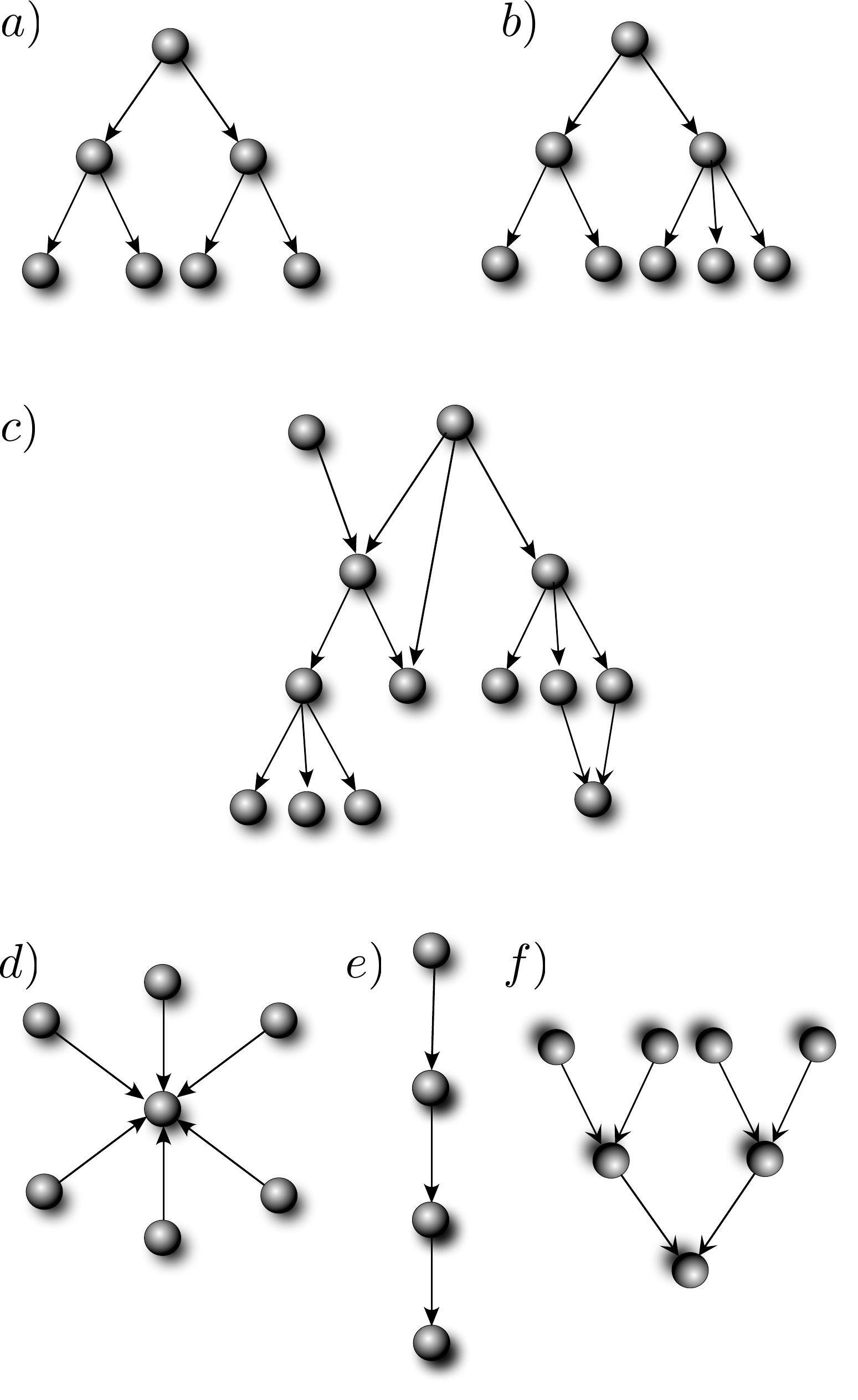}
\caption{Causal graphs intuitively capturing different degrees of hierarchy.
a)  A symmetrical tree-like causal graph showing the  ideal hierarchical structure assumed in this work 
b) An asymmetrical tree-like hierarchical graph. 
c)  A causal graph, with a  pseudo-pyramidal condition violated by
  the presence  of more  than one  maximal node, short-cuts between
   layers. Intuitively non hierarchical  and anti-hierarchical structures illustrated
 by d) an anti-hierarchical star graph with $|V-1|$ maximals exhibiting and inverted
 pyramidal condition (note that inverting the arrows we would have a tree-like graph)
e) an ordered but no hierarchical  linear chain violating the pyramidal condition and 
f) an inverted tree-like structure where no definiteness and pyramicity conditions are
 satisfied onwards but completely satisfied backwards.}
\label{fig2}
\end{center}
\end{figure}

Let us summarize the above  statements $1)$, $2)$  and $3)$ and
their consequences. The so-called {\em definiteness condition} implies
that {\em  there is  no uncertainty} in  identifying the premise  of a
given causal relation, i.e., the node that immediately governs a given
node.    Taking  into   account  the   definition  of   causal  graph,
(essentially,  a  connected  DAG)  this first  statement  restrict  to  tree structures the number of
candidates for perfectly hierarchical structures. Note that these trees  -including linear chains- have
a  single  root node,  i.e., $m=1$. The  {\em  pyramidal
  condition}  rules  out  from   the  set  of  perfectly  hierarchical
structures  those DAGs  having linear  chains in  their  structure. In
other words,  nodes  displaying $k_{in}=k_{out}=1$ are not allowed. We
observe that, according to the pyramidal condition,
\[
|\omega_i|\leq\frac{|\omega_{i+1}|}{2}.
\]
Therefore,  it is  straightforward to  conclude that  the  most simple
representation of  an ideal hierarchical structure is  a binary tree  -see fig (\ref{fig2}a)-, in  which the above
inequality becomes  equality for  all successive layers. This is 
consistent with standard graph theory and the definition of perfect 
binary tree. Finally, the
{\em symmetry  condition}, optional  for our definition  of hierarchy,
rules  out those trees  which are  not perfectly  symmetrical. Whereas trees shown in figs.  (\ref{fig2}a) and (\ref{fig2}b) can  be considered  hierarchical by
virtue of conditions $1)$ and $2)$, condition $3)$ makes a distinction
between them, being  only perfectly hierarchical
the first one.

As a  final remark, let us note  that  one can  build
non-hierarchical  and anti-hierarchical structures,  by simply 
violating some of the conditions we stated above, namely the pyramidal
condition alone  -fig. (\ref{fig2}e)-  or both the  definiteness and
the  pyramidal condition,  -see fig.  (\ref{fig2}d and \ref{fig2}f). 
It is easy to see that a quantitative estimator of hierarchy should
account for limit cases and place as intermediate point 
structures such  as  the  one  depicted in  fig.  (\ref{fig2}c).

\subsection{Topological Richness and Reversibility}

The   above   features  describe   the   {\em  perfect}   hierarchical
structure. In this section we go  further and we provide the basis for
a  definition  of a  \textit{hierarchical  index}  of  a causal  graph
grounded in the framework of information theory. This index provides a
quantitative estimation  of how far is  a given causal  graph from the
conditions of a perfect hierarchy.

In the following subsections we will define two entropies for a causal
graph,  attending to the  top-down and  bottom-up observations  of the
causal  graph  according to  the  onward  and  backward flows  in  the
graph.  The aim  of this  mathematical  formalism is  to quantify  the
impact of the number of pathways in the causal graph. Specifically, we
will  consider the  balance  between the  \textit{richness} of  causal
paths (a  top-down view)  versus the
\textit{uncertainty} when going back  reversing the causal flow (i.e.,
a bottom-up perspective). Then, attending to the direction of the flow
we interpret the  top-down view as a richness  whilst the bottom-up as
an  uncertainty  in terms  of  topological  reversibility as  recently
introduced  in \cite{Corominas-Murtra:2010}.  Arguably, the  larger is
the number of decisions  going down,  the higher  is the  richness of
causal paths. Similarly, the larger the number of alternative pathways  to climb 
up, the larger will be the uncertainty in
recovering  the causal  flow. In  the following  subsections,  we will
explore, within the framework of information theory, the relationship between
diversity and uncertainty  and the their impact in  the fulfillment of
the hierarchy conditions. We begin  this section with a brief revision
of  the  core  concepts of  information  theory,  to  be used  as  our
theoretical framework.

According   to   classical   information  theory   \cite{Shannon:1948,
  Khinchin:1957, Ash:1990, Thomas:2001}, let  us consider a system $S$
with $n$ possible  states, whose occurrences are governed  by a random
variable $X$  with an associated  probability mass function  formed by
$p_1,  ...,  p_n$.   According  to  the  standard  formalization,  the
\textit{uncertainty} or {\em entropy} associated to $X$, to be written
as $H(X)$, is:
\begin{equation}
H(X)=-\sum_{i\leq n}p_i\log p_i,
\label{DefEntrop}
\end{equation}
which is  actually an  average of $\log(1/p(X))$  among all  events of
$S$,  namely,   $H(X)=\left\langle  \log(1/p(X))\right\rangle$,  where
$\langle...\rangle$ is the {\em  expectation} or average of the random
quantity  between parentheses.   Analogously, we  can define  the {\em
  conditional  entropy}.  Given  another system  $S'$  containing $n'$
values or  choices, whose  behavior is governed  by a  random variable
$Y$,  let  $\mathbb{P}(s'_i|s_j)$ be  the  conditional probability  of
obtaining $Y=s'_i\in S'$  if we already know $X=s_j\in  S$.  Then, the
conditional entropy  of $Y$  from $X$, to  be written as  $H(Y|X)$, is
defined as:
\begin{equation}
H(Y|X)=-\sum_{j\leq    n}p_j\sum_{i\leq    n'}\mathbb{P}(s'_i|s_j)\log
\mathbb{P}(s'_i|s_j).
\label{H(Y|X)}
\end{equation}


\subsubsection{Topological reversibility: Definiteness condition}

The first task is to study the degree of reversibility of causal paths, thereby 
considering the role of the {\em definiteness condition}. This will be evaluated
 by computing the uncertainty in reversing the process starting from a given node
 in $\mu$. The formalism used in this section is close to the one developed in \cite{Corominas-Murtra:2010}.

We first proceed to define  the probability distribution from which the entropy will be evaluated. 
Accordingly, the  probability to  chose a path $\pi_k\in \Pi_{M\mu}$ from node  $v_i\in\mu$ by  making a
random decision at every crossing when {\em reverting} the causal flow is:
\begin{equation}
\mathbb{P}(\pi_k|v_i)=\prod_{v_i\in v(\pi_k)}\frac{1}{k_{in}(v_j)}.
\label{Pprod}
\end{equation}
The conditional entropy obtained when reverting the flow from $v_i\in \mu$ will
be:
\begin{eqnarray}
H({\cal G}|v_i)=-\sum_{\pi_k\in\Pi (v_i)}\mathbb{P}(\pi_k|v_i)\log\mathbb{P}(\pi_k|v_i)
\end{eqnarray}
The overall  uncertainty of ${\cal G}$, written  as $H({\cal G}|\mu)$,
is computed by averaging $H$ over all minimal nodes, i.e:
\begin{widetext}
\begin{eqnarray}
H({\cal   G}|\mu)&=&\sum_{v_i
  \in \mu}q(v_i) H({\cal G}|v_i)\nonumber\\
  &=&-\sum_{v_i    \in   \mu}q(v_i)   \sum_{\pi_k\in\Pi
  (v_i)}\mathbb{P}(\pi_k|v_i)\log\mathbb{P}(\pi_k|v_i)\nonumber\\ 
  &=& \sum_{v_i\in\mu}q(v_i)\sum_{\pi_k\in\Pi (v_i)}\left[\sum_{v_j\in v(\pi_k)}\mathbb{P}(\pi_k|v_i)\log (k_{in}(v_j))\right]\nonumber\\
  &=&\sum_{v_i\in\mu}q(v_i) \sum_{v_j\in V(\Pi(v_i)}\log (k_{in}(v_j))\left[\sum_{\pi_k:v_j\in v(\pi_k)}\mathbb{P}(\pi_k|v_i)\right]\nonumber\\
&=&\sum_{v_i\in\mu}q(v_i) \sum_{v_k\in V\setminus M}\phi_{ik}({\cal G})\log k_{in}(v_k).
\label{Entropia1}
\end{eqnarray}
\end{widetext}
where we assume, unless indicated:
\[
(\forall v_i\in \mu)\;\;q(v_i)=\frac{1}{|\mu|}
\]
Instead  of a vector,  now we  construct a  $(n-m)\times (n-m)$ matrix,
$\Phi({\cal  G})$ accounting for  the combinatorics  of paths  and how
they contribute to the computation of entropy.
\begin{eqnarray}
[\Phi({\cal    G})]_{ij}\equiv\phi_{ij}({\cal   G})=\sum_{\pi_k:v_j\in
  v(\pi_k)}\mathbb{P}(\pi_k|v_i).\nonumber
\end{eqnarray}
This represents  the probability to  reach $v_j$ starting  from $v_i$.
Now  we derive  the general  expression  for $\Phi$.   To compute  the
probability to  reach a given node,  we have to take  into account the
probability to follow a given  path containing such a node, defined in
(\ref{Pprod}).  To  rigorously connect it to the  adjacency matrix, we
first    define    an    auxiliary,   $(n-m)\times    (n-m)$    matrix
$\mathbf{B}({\cal G})$, namely:
\begin{equation}
B({\cal G})_{ij}=\frac{A_{ij}({\cal G})}{k_{in}(v_i)},
\label{B}
\end{equation}
where $v_i,v_j\in V\setminus M$.   From this definition, we obtain the
explicit dependency of $\Phi$ from the adjacency matrix, namely,
\begin{equation}
\phi_{ij}({\cal  G})=\sum_{k\leq  L({\cal G})}\left(\left[\mathbf{B}^T\right]^k({\cal
  G})\right)_{ij}.
\label{phi(B)}
\end{equation}
and accordingly, we have
\begin{equation}
\phi_{ii}({\cal G})= \left(\left[\mathbf{B}^T\right]^0({\cal G})\right)_{ii}=1.
\end{equation}
Therefore, we already obtained the explicit form of such a conditional
entropy, namely:
\begin{equation}
H({\cal    G}|\mu)=\sum_{v_i\in    \mu}q(v_i)\sum_{v_k\in   V\setminus
  M}\phi_{ik}({\cal                                             G})\cdot\log
k_{in}(v_k).
\end{equation}
Assuming equiprobability, the above expression leads to:
\begin{equation}
H({\cal      G}|\mu)=\frac{1}{|\mu|}\sum_{v_i\in      \mu}\sum_{v_k\in V\setminus M}\phi_{ik}({\cal G})\cdot\log k_{in}(v_k).
\label{HGeneral}
\end{equation}

\subsubsection{Topological richness: Pyramidal condition}

Let us now estimate the topological richness of a causal graph, i.e., the average amount of information 
needed to describe a given top-down path within the structure. Let us observe that the kind of question 
we are trying to answer is the same than the one explored above, but considering the top-down approach. 
Therefore, the mathematical form of this quantity, to be referred as $H(G|M)$, will be formally analogous
 to the previous one, but considering that we are going onwards according to the causal flow. Thus:
\begin{equation}
H(G|M)=\frac{1}{m}\sum_{v_i\in M}\sum_{v_k\in V\setminus \mu}\psi_{ik}({\cal G})\cdot\log k_{out}(v_k),
\end{equation}
where $m$ is the cardinality of $M$ -the set of maximal nodes. $\psi_{ik}$ is analogous to $\phi_{ik}$ of equation (\ref{HGeneral}). In this case, elements $ik$ of matrix $\Psi$ represent the probability to cross node $v_k$ departing from $v_i\in M$ according to the causal flow. The explicit expression of $\Psi$ is defined from matrix $\mathbf{B}'({\cal G})$:
\[
B'({\cal G})_{ij}=\frac{A_{ij}({\cal G})}{k_{out}(v_i)},
\]
Then,
\begin{equation}
\psi_{ij}({\cal  G})=\sum_{k\leq  L({\cal G})}\left(\left[\mathbf{B}'\right]^k({\cal G})\right)_{ij}.
\label{phi(B)}
\end{equation}
and as above, we have
\begin{equation}
\psi_{ii}({\cal G})= \left(\left[\mathbf{B}'\right]^0({\cal G})\right)_{ii}=1.
\end{equation}

\subsection{Hierarchy}

As we shall see, the above definition of information will bring us the
ingredients  to define  a hierarchy  index  according to  the list 
detailed  in  section   \ref{section:Conceptual}.   Roughly
speaking, what  we propose in the  following lines is  to evaluate the
balance  between the  pyramidal  structure of  the  graph against  the
degree of  reversibility of the  paths it generates, i.e,  the balance
between $H({\cal  G}|M)$ and $H({\cal  G}|\mu)$. However, in  order to
rigorously characterize hierarchy, we  need to properly treat the
studied  graph attending to the different layers of its feedforward structure. The analysis of the graph structure allows
us to  identify and  quantify deviations  from the
perfect structure at  any level of the graph.  The starting
point will involve the characterization of a layered structure within
the graph defining a partition $W$ of the set of nodes.

\subsubsection{Dissecting the layer structure}

\begin{figure}
\begin{center}
\includegraphics[width=7.5cm]{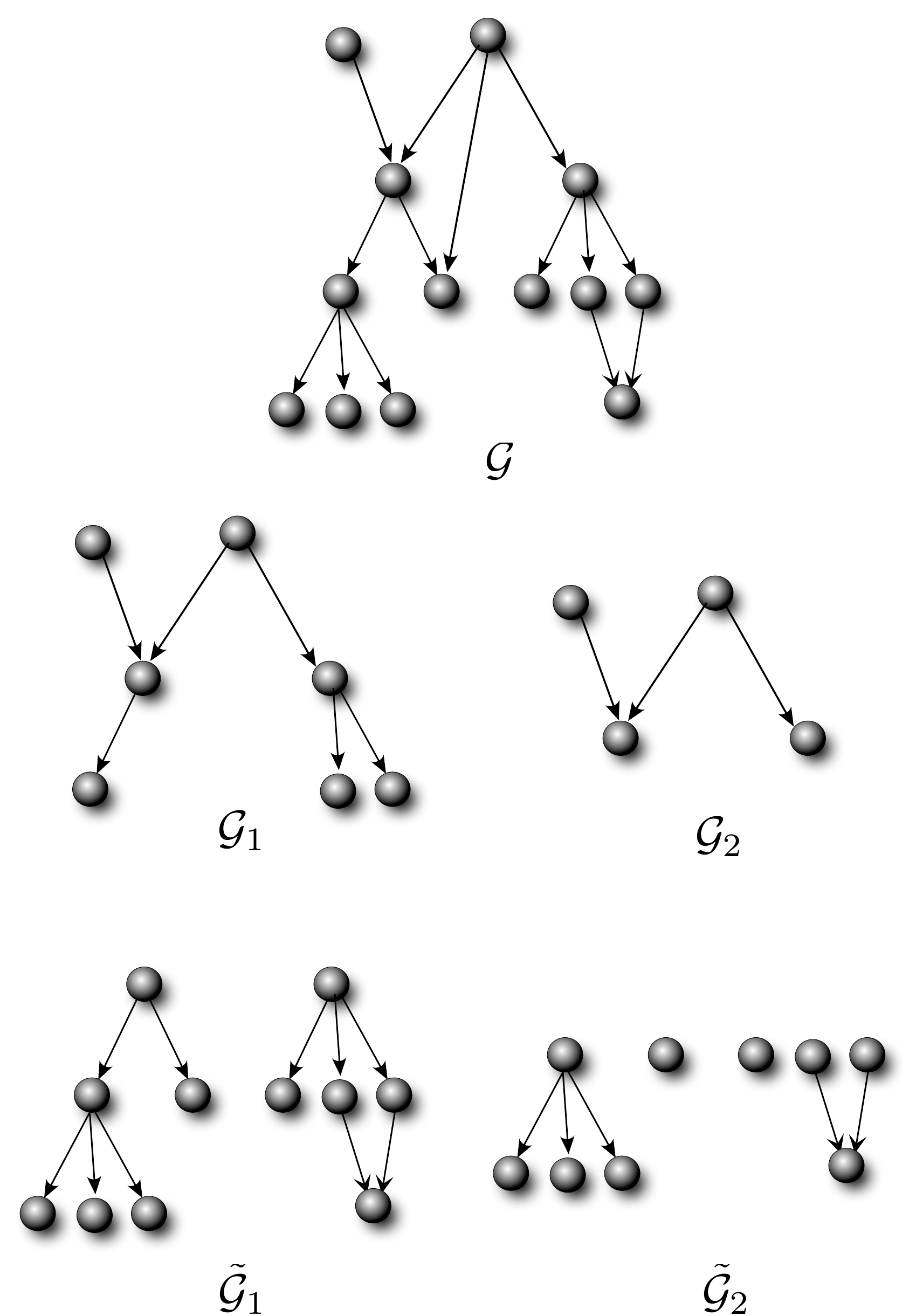}
\caption{How to obtain the different  subgraphs   ${\cal  G},{\cal
    G}_1,{\cal G}_2,\tilde{\cal G}_1,\tilde{\cal G}_2$ involved in the
  evaluation of hierarchy. ${\cal  G}_1$ and ${\cal G}_2$ are obtained
  through successive application  of bottom-up leaf removal algorithm,
  which    implies     that    we    remove     all    nodes    having
  $k_{out}=0$.  $\tilde{{\cal   G}}_1$  and  $\tilde{\cal   G}_2$  are
  obtained by  the successive application  of a top-down  leaf removal
  algorithm, thereby removing the  nodes having $k_{in}=0$. We observe
  that the  generation of $\tilde{{\cal G}}_1$  and $\tilde{\cal G}_2$
  implies the breaking of the net -see text.}
\label{fig:LeafRemovals}
\end{center}
\end{figure}
Given the DAG ${\cal G}(V,E)$, let  us define two  partitions of $V$,
$W=\{\omega_1,...,\omega_m\}$                                       and
$\tilde{W}=\{\tilde{\omega}_1,...,\tilde{\omega}_m\}$.  The members of
such partitions are the layers of the DAG by either performing a top down
or bottom up {\em leaf removal algorithm} \cite{Lagomarsino2007, Rodriguez-Caso2009}. 
Specifically,  the first members of such partitions are defined as:
\[
\omega_1=\{v_i\in V:k_{out}(v_i)=0\}
\]
and
\[
\tilde{\omega}_1=\{v_i\in V:k_{in}(v_i)=0\}.
\]
We observe that $\omega_1=\mu$ and that $\tilde{\omega}_1=M$.
With  the  above  subsets of  $V$  we  can  define the  graphs  ${\cal
  G}_1(V_1, E_1)$, and $\tilde{\cal G}_1(\tilde{V}_1, \tilde{E}_1)$ in
the following way:
\[
V_1=V\setminus\omega_1;\;\;E_1=E\setminus \{\langle v_i,v_k\rangle:v_k\in \omega_1\}.
\]
and
\[
\tilde{V}_1=V\setminus\tilde{\omega}_1;\;\;\tilde{E}_1=E\setminus
\{\langle v_i,v_k\rangle:v_k\in \tilde{\omega}_1\} .
\]
respectively.  Similarly, we build $\omega_2,...,\omega_{|W|}$ as:
\begin{eqnarray}
\omega_2&=&\{v_i\in
V_1:k_{out}(v_i)=0\}\nonumber\\                 \omega_{|W|-2}&=&\{v_i\in
V_{|W|-1}:k_{out}(v_i)=0\}\nonumber
\end{eqnarray}
and
\begin{eqnarray}
\tilde{\omega}_2&=&\{v_i\in
\tilde{V}_1:k_{in}(v_i)=0\}\nonumber\\ \tilde{\omega}_{|W|}&=&\{v_i\in
\tilde{V}_{|W|-2}:k_{in}(v_i)=0\},\nonumber
\end{eqnarray}
respectively.  We  therefore defined  two  sequences  of subgraphs  of
${\cal G}$ ordered by inclusion, namely:
\[
 {\cal G}_{|W|-1}\subseteq  ...\subseteq {\cal
  G}_1\subseteq {\cal G}.
\]
and
\[
 \tilde{\cal    G}_{|W|-1}\subseteq
...\subseteq \tilde{\cal G}_1\subseteq {\cal G}.
\]
where it is easy to observe that:
\[
{\cal G}_{|W|-1}=M\;\;{\rm and}\;\;\tilde{\cal G}_{|W|-1}=\mu.
\]
In fig. (\ref{fig:LeafRemovals}) we describe the generation of 
these subsets of graphs for a given toy model of DAG. 
In summary, we  constructed two collections of subsets  by finding the
layered structure  using a  bottom-up leaf removal  algorithm (pruning
the  elements having  $k_{out}=0$  successively) and  a top-down  leaf
removal  algorithm  (i.e., pruning  the  elements having  $k_{in}=0$).
Notice that, even if
 \[
|W|=|\tilde{W}|,
\]
one cannot assume
\[
w_i= \tilde{w}_{|W|-i},
\]
except in symmetrical cases.

\subsubsection{The hierarchy index}

\begin{figure*}
\begin{center}
\includegraphics[width=18cm]{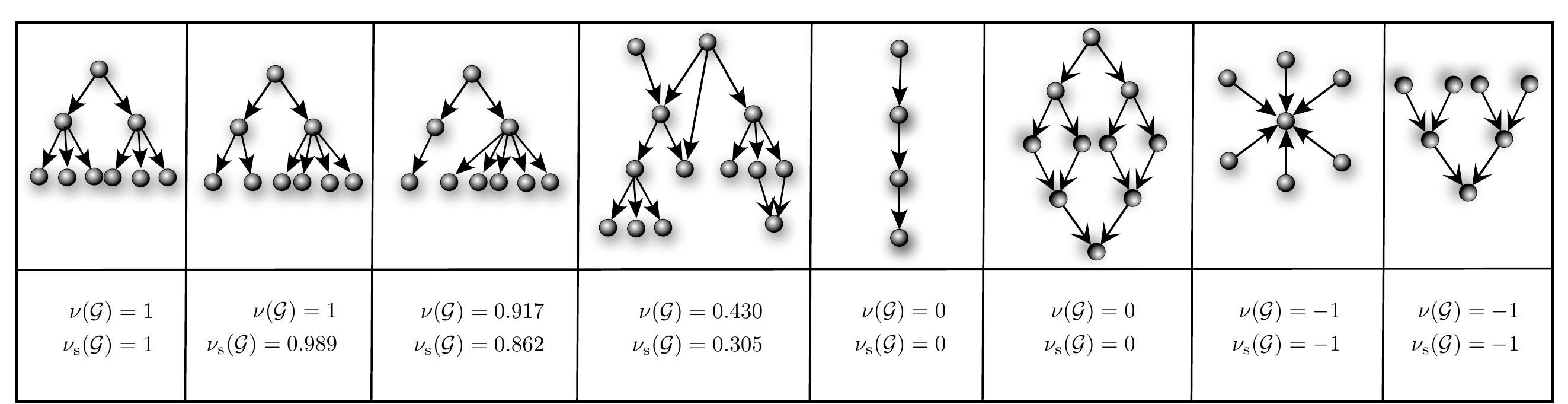}
\caption{Different values of the hierarchy index corresponding to some toy DAGs. $\nu({\cal G})$ refers to the hierarchical index where the symmetry is not considered and $\nu_{\rm s}({\cal G})$ refers to the hierarchical index where symmetry is taken into account -see text. }
\label{FigGrafets}
\end{center}
\end{figure*}

In order to generate a normalized estimator $f({\cal G})$ (between $-1$ and $1$) accounting for the balance between 
$H({\cal G}|M)$ and $H({\cal G}|\mu)$ we will define it as:
\[
f({\cal G})\equiv \frac{H({\cal G}|M)-H({\cal G}|\mu)}{\max\{ H({\cal G}|M),H({\cal G}|\mu)\}}.
\]
Since both the layered  structure and  its pyramidal  composition  must be
taken  into account,  the  hierarchical  index of  the  graph must  be
weighted taking into account the  successive layers of the system. This
avoids to identify as completely hierarchical those structures not perfectly satisfying
the {\em pyramidal condition}.  Therefore, the {\em hierarchical index
  of a feed-forward net}, to be indicated as $\nu({\cal G})$ will be the average among the $|W|-2$ subgraphs ${\cal G}_1,...,{\cal G}_k,...$, the $|W|-2$ subgraphs $\tilde{\cal G}_1,...,\tilde{\cal G}_k,...$ and ${\cal G}$ itself -note that we average between $2|W|-3$ objects, i.e.:
\begin{eqnarray}
\nu({\cal G})=\frac{1}{2|W|-3}\left(f({\cal G})+\sum_{i<  |W|-1}f({\cal  G}_i)+f(\tilde{\cal G}_i)\right).
\label{hierarchy}
\end{eqnarray}

It is strictly necessary to take into account all these subgraphs in order to identify any violation of the hierarchy conditions at any level of the structure.
We can go a step further by imposing symmetry in the pyramidal structure as suggested above, to
distinguish among different topologies such as those displayed in figure (\ref{fig2}a). Let us indicate by 
$\Pi_{M\mu}({\cal  G}_k)$ the  set of  paths from  $M$ to
$\mu$ present in  the graph ${\cal G}_k$. The  so-called {\em Jensen's
  inequality} \cite{Thomas:2001} provides  an upper  bound  of  the information  content
which, in our case reads:
\begin{eqnarray}
{\rm a)}&&H({\cal G}_k|M)\leq \log |\Pi_{M\mu}({\cal G}_k)|,\nonumber\\
{\rm b)}&&H({\cal G}_k|\mu)\leq \log |\Pi_{M\mu}({\cal G}_k)|.\nonumber
\end{eqnarray}
We observe that a) is only achieved when all $|\Pi_{M\mu}|$ from $M$ to $\mu$ are equiprobable, 
being this equiprobability an indicator of symmetry. The same applies to b), but now 
we consider the bottom-up estimator, when paths are considered from $\mu$ to $M$. 
In this case, attending to  the symmetric condition we can  define and estimator analogous to  $f$, namely, 
\[
g({\cal G})=\frac{H({\cal G}_k|M)-H({\cal G}_k|\mu)}{\log|\Pi_{M\mu}({\cal G})|}
\]
Accordingly, the symmetrized version of the hierarchy index, $\nu_s({\cal G})$, would be:
\begin{eqnarray}
\nu_s({\cal G})=\frac{1}{2|W|-3}\left(g({\cal G})+\sum_{i< |W|-1} g({\cal G}_i)+g({\cal G}_i)\right).
\label{hierarchySym}
\end{eqnarray}

To ensure the  consistency of $\nu$ and $\nu_s$ we must 
overcome  the last  conceptual problems. When  performing the
leaf  removal operation  one can  break  the graph  in many  connected
components  having no connections  among them.  Let us indicate as 
${\cal C}_1(i),...,{\cal C}_k(i)$ the set of components of our graph ${\cal  G}_i$, each of them 
obtained from the  set $V_k(i)\subseteq  V_i$ of  nodes and their. In this case, the natural way to proceed is to average the
individual  contributions  of the  different  connected components  of
${\cal G}_i$  or $\tilde{\cal G}_i$  according to the number  of nodes
they have against $|V_i|$ or $\tilde{V}_i$, leading to:
\[
f({\cal G}_i)\equiv\frac{1}{|V_i|} \sum_{{\cal C}_k(i)}|V_k(i)|f({\cal C}_k(i)).
\]
The same applies for $g({\cal  G}_i)$ for the computation of
$\nu_s$.  Finally,  we  impose,  for  both  mathematical  and
conceptual consistency that:
\[
(\max\{H({\cal G}|M),H({\cal G}|\mu)\}=0)\Rightarrow (\nu({\cal G})\equiv 0).
\]
Furthermore,   if
$E=\varnothing$, (i.e., the case where  the graph consists of a single
node):
\[
\nu({\cal G})\equiv 0.
\]
According to this formulation  some scenarios would lead to $\nu({\cal
  G})=0$.  The  simplest one  is  the  just  mentioned by  definition,
consisting of  a single node.  Another one is the  linear feed-forward
chain  having $2$  or more  linked nodes.  It is  clear that  in these
cases, $H({\cal  G}|M)=H({\cal G|\mu})=0$. It is worth  to stress that
this particular  situation matches  with the causal  graph of  a total
order relation, and  therefore, we have the way  to differentiate this
particular graph from other structures having null hierarchy. Finally,
a third class of structures  belongs to the family of non-hierarchical
graphs.  They  give  $\nu({\cal  G})=0$ since  $H({\cal  G}|M)=H({\cal
  G}|\mu)$.  This is the
case  of Erd\"os  R\'enyi DAGs  or DAG  cliques. In  these  cases, the
causal graph  is not hierarchical  because all the diversity  of paths
generated when  crossing the causal  flow downwards is  neutralized by
the uncertainty in recovering any causal path backwards.

\subsection{Numerical Exploration}
\begin{figure}
\begin{center}
\includegraphics[width=8.5cm]{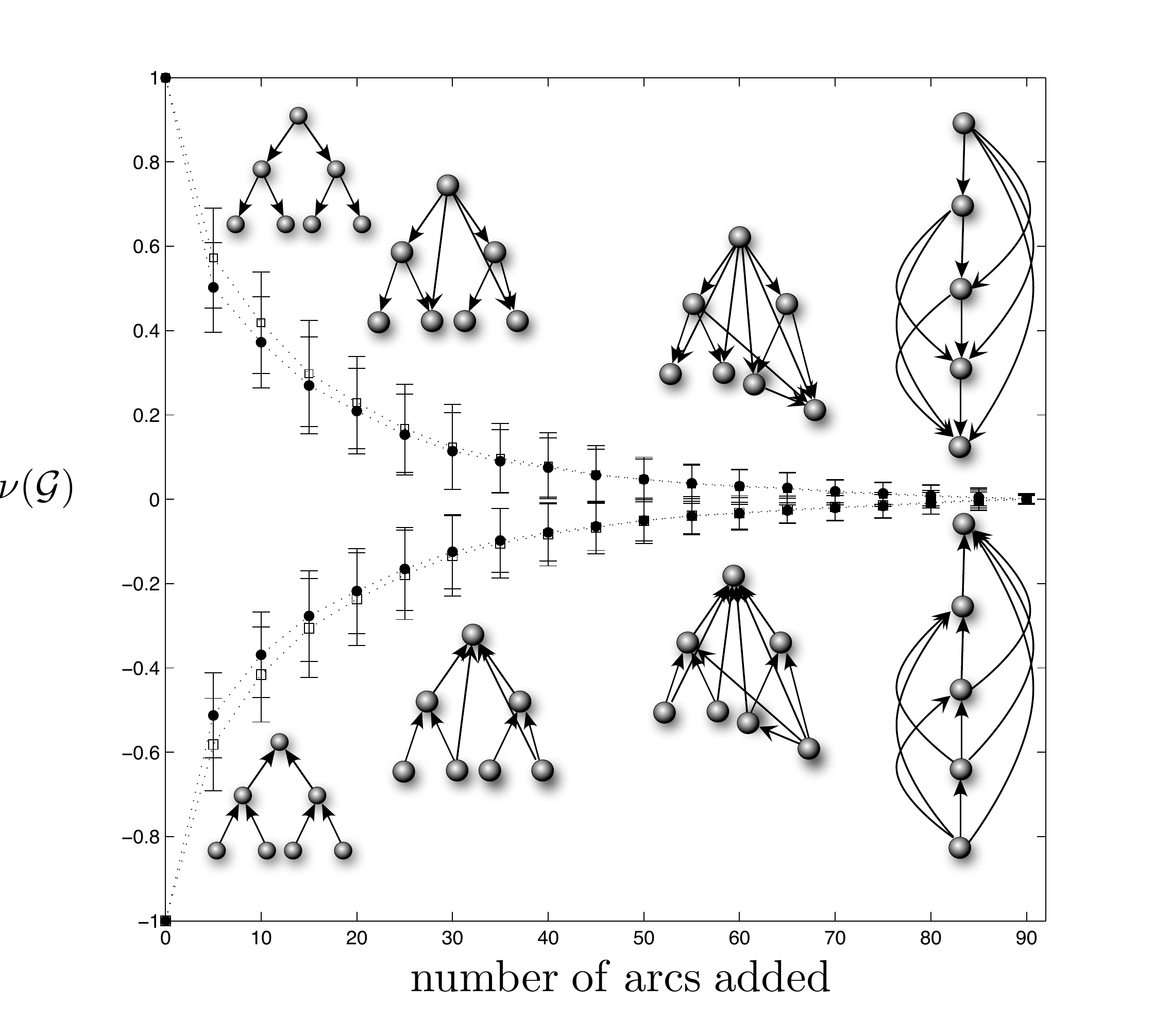}
\caption{The evolution of the hierarchy  index. We start with a binary
  tree and a binary inverted tree with $n=15$ and $|E|=14$. 
For  both graphs  we add,  in an  iterative way,  arcs at
  random until  we reach the feed forward  clique configuration.  Note
  that  the link  addition  to the  two  initial tree-like  structures
  converge  in the  same full  connected configuration  which contains
  ${15 \choose  2}$ links. For both experiments
  squares  represent $\nu$  values, and  triangles $\nu_{\rm  s}$.  As
  expected, $|\nu|>|\nu_{\rm s}|$, except in limit cases.  The small graphs provide 
a  visual clue for type  of
  structures we are obtaining  through the arc enrichment process. For
  every  point  in  the   chart,  mean  and  standard  deviation  were
  calculated from $250$ replicas.  Entropies were computed considering
  $\log_2$.}
\label{FigEvolution}
\end{center}
\end{figure}

In this section we evaluated the hierarchy of several toy models in order to intuitively grasp the scope of the measure. 
In fig. (\ref{FigGrafets}) we evaluated the hierarchy index (both the raw one and the symmetrical one) for several structures leading to hierarchical, anti-hierarchical and non-hierarchical structures. The figure illustrates the impact of number of maximals and minimals and the multiplicity of pathways in relation to the existence of a pyramidal and predictable structure. We observe that deviations from tree and inverted tree configurations lead to  a non binary interpretation of hierarchy. 
 
Furthermore, we measured (fig. (\ref{FigEvolution}) the impact in terms of hierarchy of arc addition 
preserving the acyclic character. Staring from two  extreme tree graphs (the feedforward and the inverted ones, respectively) 
we add arcs at random until we reach a fully connected feed-forward structure in both situations.
 We consider the starting point of our numerical experiment a binary tree, ${\cal T} (V, E)$ containing $n=15$ nodes. We construct an  inverted binary tree ${\cal T}' (V, E)$ by the transposition of the adjacency  matrix of ${\cal T} (V, E)$.
 In both graphs we say  that, consistently with the ordering property of DAGs,  given  an arc  $(\langle v_i,v_j\rangle \in E)$ then $(i<j)$.
In an iterative process we construct two  new DAGs ${\cal G}^i(V, E^i)$ where $i$ labels
the number of  additions of new arcs to the underlying ${\cal T} (V, E)$ and ${\cal T}' (V, E)$.
The process ends when graphs achieve the directed acyclic clique condition, i.e., the linearly ordered  graph ${\cal G}^*=(V,E^*)$ containing $15$ nodes:
\[
(\forall v_i, v_j\in V):(i>j)(\langle v_j, v_i\rangle\in E^*)
\]
For statistical significance, we performed $250$ replicas of the numerical experiment. Then $\langle\nu\rangle$ and $\langle \nu_{\rm s}\rangle$ 
and their respective standard deviations were calculated for each set of iterations. Fig. (\ref{FigEvolution}) shows
that starting from an initial value of  $\nu=1$ for  ${\cal T} (V, E)$ and $\nu=-1$ for ${\cal T}' (V, E)$, the addition of feed-forward arcs causes a decrease in absolute value of the hierarchical indexes 
until  $\nu=\nu_{\rm s}=0$ corresponding to a total linear ordered structure where every
possible feed-forward path is included in the graph. As expected, $|\nu|>|\nu_{\rm s}|$ except in the extreme full-connected cases. 

\begin{figure}
\begin{center}
\includegraphics[width=8cm]{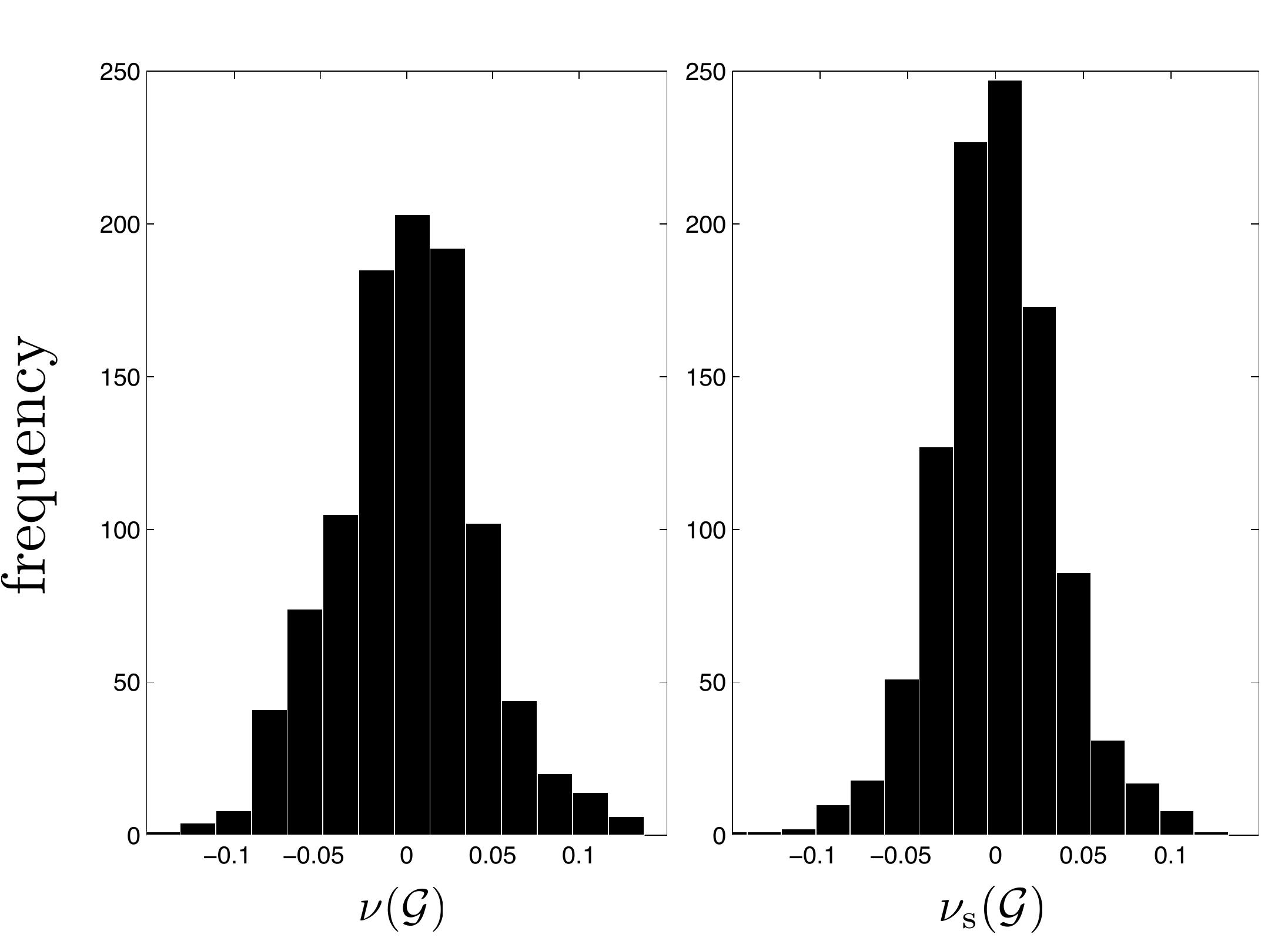}
\caption{Distribution of $\nu$ and $\nu_{\rm s}$ for an ensemble of $1,000$ replicas of directed acyclic ER graphs of $|V|=500$ and $\langle k \rangle=4$ caption. Numerical results show Gaussian-like distributions centered at zero. Notice that  $\nu_{\rm s}$ distribution displays a narrower variation than the $\nu$ one in agreement to the $|\nu|\geq|\nu_{\rm s}|$ inequality.}
\label{ER}
\end{center}
\end{figure}

We finally test the case of a directed acyclic Erd\"os R\'enyi (ER) graph  ${\cal R}(V,E)$. This is an interesting example of a topologically homogeneous DAG with non-correlation in terms of $k_{in}$ and $k_{out}$ \cite{Goni:2010}. Graphs were obtained by the construction of an undirected ER graph ${\cal G}_{ER}(V,E^u)$ where $E^u$ is the set of edges (undirected links). Directed acyclic condition was obtained by a process of random numbering of nodes \cite{Goni:2010}. The direction of the arrows was defined attending 
\[
\langle v_i, v_j\rangle \in E: (\{v_i,v_j\} \in E^u  \wedge i<j)
\]
(i.e., condition depicted in eq. (6))
Fig. \ref{ER} shows a representative behavior of the null hierarchical  character of  ${\cal R}(V,E)$ ensembles. Notice that the normal distribution is centered for $\nu$ and $\nu_{\rm s}$ at zero values, indicating that such random structures have not any hierarchical organization. 
%

\section{Discussion}

Hierarchical  patterns  are  known  to  pervade  multiple  aspects  of
complexity. In spite of their  relevance, it is not obvious in general
how to  formalize them in terms  of a quantitative  theory. This paper
presents a definition of hierarchy to be applied to the so-called {\em
  causal graphs}, i.e., connected,  directed acyclic graphs where arcs
depict some  direct causal relation between the  elements defining the
nodes. It  is therefore a measure  of hierarchy over  the structure of
the causal flow. The conceptual basis of this measure is rooted in two
fundamental features  defining hierarchy: the absence  of ambiguity in
recovering  the   causal  flow  and   the  presence  of   a  pyramidal
structure. The  hierarchy index presented here  weights the deviations
from such  general properties. The specific expression  for this index
is derived  using techniques and concepts from  information theory. It
is shown, thus,  that the requirements of hierarchy  naturally fit the
tension between  richness in causal  paths against the  uncertainty in
recovering  them  depicted  by   a  balance  between  two  conditional
entropies.

Under our previous assumptions, we have shown  that the
feed-forward tree is the structure that fully satisfies the conditions
for a perfect hierarchical system.  Interestingly, trees  as perfect
representations    of   hierarchies    is    a   long-standing    idea
\cite{Whyte1969}.   In  this   way,  our   mathematical  formalization
establishes a bridge between the qualitative idea of hierarchy and its
quantification. Our  approach allows to  measure the hierarchy  of any
system provided  that it  can be represented  in a  feedforward causal
graph.

Throughout the  paper we emphasized that although  hierarchy is deeply
tied to  order, there  are strong  reasons to go  beyond it.  The most
obvious  one is that  {\em order}  is a  well established  concept and
therefore, there  is no need  for the use  of a different word,  if we
identify it with hierarchy. But  another important issue must be taken
into account tied to the  intuitive notion of hierarchy we propose. We
propose that hierarchy  must  feature the  pyramidal  nature of  the
connective  patterns and  that  this pyramidal  structure  must be  in
agreement  with  the  top-down  nature  of the  feed-forward  flow  of
causality.

Information theory reveals extremely  suitable to define a hallmark to
study hierarchy in  the terms described in this  paper: the richer the
structure (but  at the same  time, reversible, in  topological terms),
the more  hierarchical it  is.  In this  way, since the  conditions we
defined for a system to  be perfectly hierarchical lead us to conclude
that a feed  forward tree is the perfect  hierarchical structure since
maximizes the richness without loss  of predictability. It is worth to
note  that precisely,  the pyramidal  condition  is the  key point  to
guarantee the the predictability.  The extreme case is the feedforward
clique  -  see  fig.   \ref{FigGrafets}).  Although  richness  can  be
increased through pathway redundancy, this effect cancels out due to 
decreased predictability, leading to a non  hierarchical structure. A
particular case  is the linear  chain. This representation of  a total
order relation  has null values of  both entropies. In  other words, a
perfect predictable system  but without richness. It is  worth to note
that   both  cases   are  not   pyramidal  structures.   By  contrast,
anti-hierarchical ones exhibit an inverted pyramidal structure leading
to  a different  effect. From  the perspective  of our  formalism, the
anti-hierarchical  organization  occurs  by  minimizing  richness  and
predictability.  In  consequence, other  structure  different than  an
inverted tree will be less  anti-hierarchical. Therefore it is easy to
see  that  the  hierarchical  index  in absolute  value  measures  the
closeness of  an (anti)-hierarchical tree  structure capturing somehow
the path complexity of the structure.

Further  work   should  explore  the  relation   of  this  theoretical
achievement   within   the   framework   of  a   formal   measure   of
complexity. Additionally,  this research could  be expanded to  a more
general class of directed  graphs containing cycles. This latter point
would  be achieved by  properly defining  a measure  of {\em  how well
  ordered} is a net, for it  is clear that the presence of cycles will
generate conceptual problems in the identification of the causal flow.

\begin{acknowledgments}
This  work  was  supported   by  the  EU  $6^{th}$  framework  project
ComplexDis (NEST-043241, CRC and JG), the UTE project CIMA (JG), Fundaci\'on Marcelino Bot\'in (CRC), the James
McDonnell Foundation  (BCM and  RVS) and the Santa  Fe Institute  (RVS). We
thank Complex System Lab members for fruitful conversations and an anonymous reviewer for his/her constructive suggestions.
\end{acknowledgments}

\end{document}